\documentclass[aps,twocolumn,showpacs]{revtex4}
\usepackage[dvips]{graphicx}
\usepackage{amsmath}
\newcommand{\bea}{\begin{eqnarray}}
\newcommand{\eea}{\end{eqnarray}}
\newcommand{\be}{\begin{equation}}
\newcommand{\ee}{\end{equation}}

\newcommand{\nn}{\nonumber}

\newcommand{\uG}{\underline{G}}
\newcommand{\uT}{\underline{T}}

\newcommand{\uK}{\underline{K}}

\newcommand{\udK}{\underline{\delta K}}
\newcommand{\dK}{{\delta K}}
\newcommand{\dA}{{\delta A}}
\def\k{{\bf k}}
\def\br{{\bf r}}
\def\R{{\bf R}}
\def\p{{\bf p}}
\def\A{{\bf A}}
\def\J{{\bf J}}
\def\q{{\bf q}}
\def\v{{\bf v}}
\def\l{{\bf l}}
\def\B{{\bf B}}
\def\F{{\bf F}}
\begin{document}

\title{Singular current response from isolated impurities in
$d$-wave superconductors}

\author{Shan-Wen Tsai$^1$ and P. J. Hirschfeld$^{1,2}$}

\affiliation{$^1$ Department of Physics, University of Florida, PO
Box 118440, Gainesville FL 32611\\$^2$ Center for Electronic
Correlations and Magnetism, EP6, Univ.  Augsburg, Augsburg Germany
}
\date{\today}

\begin{abstract}
The current response of a $d$-wave superconductor containing a
single impurity is calculated and shown to be singular in the
low-temperature limit, leading in the case of strong scattering to
a $1/T$ term in the penetration depth $\lambda(T)$ similar to that
induced by Andreev surface bound states.  For a small number of
such impurities,  we argue this low-$T$ upturn could be observable
in cuprate superconductors.
\end{abstract}

\maketitle

{\it Introduction.}  The quasiparticle excitations near the line nodes in the
$d$-wave superconducting order parameter of the cuprate high-$T_c$ materials
lead to well-known singularities in the current response of such systems,
resulting in nontrivial magnetic field dependences of the London penetration
depth in the Meissner state\cite{yipsauls}, and of the specific heat in the
Abrikosov vortex state\cite{volovik}. These excitations are also responsible
for the marginal thermodynamic stability of the $d$-wave state itself, which
has been related\cite{schopohl,lietalcomment}
to the famous
linear-$T$ temperature dependence of the penetration depth
$\lambda(T)$\cite{hardy}.

 Recently, it has been
pointed out by Walter et al.\cite{walteretal} and Barash et
al.\cite{barashetal}, that, in addition to  effects arising from
extended quasiparticle states alone, there are  $1/T$
contributions to the penetration depth below a crossover
temperature $T_{m0}\simeq \sqrt{\xi_0/\lambda_0}T_c$, where
$\xi_0$ is the coherence length and $T_c$ the critical
temperature, due to
 zero-energy surface bound states.  These  states arise
 within a semiclassical picture when a quasiparticle reflected
 from the surface experiences a change of sign in the order
 parameter; the
upturns therefore only occur when the surface normal of the sample
makes an angle close to $\pi/4$  with the crystal axes and hence
the antinodal directions of the $d$-wave order parameter.  A
$\lambda(T)\sim 1/T$ term in the penetration depth has in fact
been reported by Carrington et al.\cite{prozorov}, 
and attributed to the paramagnetic current carried by such
states which dominates at low $T$.  The field and interface angle
dependence also appear to be in rough agreement with the
predictions of Barash et al.\cite{barashetal}.  Upturns correlated
with disorder introduced by Zn atoms observed by Bonn et
al.\cite{bonnetal}
 seem
to vary from sample to sample, and are not currently understood.

Most works in this area have treated disorder within effective-medium
approximations which predict broadening of low-energy quasiparticle states by a
residual rate $\gamma$, which is also roughly proportional to the residual
density of states at zero energy $N(0)$.  Impurities are thus assumed to
``smear out" the nodes of the gap, and therefore inevitably cut off
the singular behavior. For example, in the work of Barash et
al.\cite{barashetal}, no upturn in $\lambda(T)$ at low $T$ is observed if
$\gamma>T_{m0}$. In the impurity-dominated regime, a quadratic temperature
dependence is generally to be expected\cite{grossetal}.  Impurity physics
is therefore thought to compete\cite{lietalcomment} with nonlocal\cite{kosztin} and
nonlinear\cite{yipsauls} effects in the Meissner state, and $\lambda(T)$
changes from linear to quadratic below whichever scale is largest.

There may be reasons, however, to doubt results obtained for the
penetration depth using the self-consistent $T$-matrix
approximation (SCTMA), which  defines a  translationally invariant
effective medium for the impurity-averaged dirty $d$-wave
superconductor.  First, it has been shown to break down in two
dimensions\cite{tsvelik}  and corrections due to weak localization
and correlated order parameter response have been under intense
investigation recently.\cite{pjhjltp}  Secondly, as with any
effective medium theory, it must break down when impurities are
sufficiently isolated.  Finally, there are clear indications that
single impurity bound states in unconventional superconductors are
analogous in some respects to surface Andreev states.\cite{goldbart}  Here 
we point out that under some
circumstances isolated strong impurities can themselves make
singular contributions to the penetration depth and other
thermodynamic quantities.  While we can present only estimates for
the effect of a thermodynamically finite density of dilute
impurities, we perform an exact calculation for the current
response of a $d$-wave superconductor in the presence of a single
strong scatterer. 

 {\it Current response for single
impurity.}  We consider a pure $d$ wave superconductor described by  Nambu
propagator $\uG^0_{\k}(\omega)=(i\omega_n \tau_0 +\xi_\k \tau_3
+\Delta_\k\tau_1)/D_\k$, where $D_\k\equiv \omega_n^2 + \xi_\k^2+\Delta_\k^2$,
$\xi_\k\equiv \epsilon_\k-\mu$ is the 1-electron spectrum measured relative to
the Fermi level, and the $\tau_i$ are Pauli matrices. The $d$-wave order
parameter on the model cylindrical Fermi surface parametrized by angle $\phi$
is $\Delta_\k=\Delta_0\cos 2\phi$. In the system with one single
$\delta$-function impurity of strength $V_0$ located at position $\R$, the
Green's function is exactly given by \bea\label{fullG}
\uG_{\k\k'}(\omega)=\uG^0_{\k}(\omega)\delta_{\k\k'} +
e^{i(\k-\k')\cdot \R}
\uG^0_{\k}(\omega)\uT(\omega)\uG^0_{\k'}(\omega), \eea 
where $\uT$ is the
exact $T$-matrix for the impurity. We now ask what the {\it change} in the
current response due to a single impurity is in this system. Note that we
intend to study at first only the {\it linear} response to an external vector
potential $\A$ \`a la Kubo, but in the ``unperturbed state" described by the
{\it exact} one-particle eigenstates with the single impurity present. This is
simply the usual nonlocal expression
for the current response
in Coulomb gauge, 
\bea
\J(\br)=\int d\br' \uK
(\br,\br')\A(\br'),
\label{response}\eea 
with
$\uK(\br,\br') = - c/(4\pi\lambda^2(T))\delta(\br-\br') + 
\udK(\br,\br')$.
Here we take
$\lambda(T)$ to be the unperturbed London penetration depth of the $d$-wave
superconductor, with pure behavior $\lambda(T)\simeq \lambda_0(1+(\log 2)
T/\Delta_0)$ for $T\ll \Delta_0$ and $\lambda_0\equiv \sqrt{mc^2/4\pi ne^2}$.
The Fourier transform $\udK
(\p,\q)$ is now the change in response due to the impurity, which may
be easily expressed in terms of the exact 1-impurity $T$-matrix using
(\ref{fullG}),

\bea && \dK^{\alpha\beta}(\p,\q)=-{e^2\over c{\cal V}} e^{i{(\p-\q)\cdot \R}}
\,T\sum_\omega\sum_\l {1\over 2} {\rm Tr} \label{dKmicro}\\&& \left\{
\v_{\l-{\q\over 2}}^\alpha\v_{\l-{\p\over 2}}^\beta\,\uG^0_{\l}(\omega) \uG^0_{\l-\q}(\omega) \,\uT(\omega) \uG^0_{\l-\p}(\omega) 
\right.\nn\\
&&+\v_{\l+{\q\over 2}}^\alpha\v_{\l+{\p\over 2}}^\beta\,\uG^0_{\l+\p}(\omega) \,\uT(\omega) 
\uG^0_{\l+\q}(\omega) \uG^0_{\l}(\omega) +{1\over {\cal V}}
\sum_\k\nn\\
&& \left.\v_{\l+{\q\over 2}}^\alpha\v_{\k-{\p\over 2}}^\beta\uG^0_{\k}(\omega) \,\uT(\omega) 
\uG^0_{\l+\q}(\omega) 
\uG^0_{\l}(\omega) \,\uT(\omega)  \uG^0_{\k-\p}(\omega) 
\right\},\nn\eea where $\omega$ is an internal fermion
Matsubara frequency, ${\cal V}$ is the volume of the system, and
$\v_\k\equiv \partial \epsilon/\partial \k$ is the electron
velocity.

We now need to solve (\ref{response}) together with Maxwell's equations to
determine the spatial dependence of the vector potential caused by the combined
effects of Meissner screening and impurity scattering. 
 Since we are primarily interested in long-wavelength effects, the perturbation
may be considered to be of order $1/N$, where $N$ is the number of atoms in the
crystal; we may therefore clearly treat the problem perturbatively, by writing
$\A(\br)=\A_0(z) +\delta \A(\br)$, where the unperturbed solution is taken to
be the London result (we assume $\lambda_0>>\xi_0$ for the homogeneous system)
$\A_0(z)=\A_0(0)\exp-|z|/\lambda_0$.  Note $z>0$ is the coordinate describing
the theorist's half-space of superconducting material, and the solution
$A(\br)$ is extended to  unphysical values $z<0$ as an  ``image vector
potential" to allow a solution by Fourier transform.
Specular
scattering of quasiparticles from the actual surface is assumed throughout.

To linear order in $\delta \A$, the problem can be cast as a
 differential rather than integro-differential equation,
\bea \nabla^2 \delta \A -{1\over \lambda(T)^2 }\delta\A =
{4\pi\over c} \int d\br' \udK (\br,\br') \A_0(z') \eea whose
solution can be obtained by Fourier transform, \bea \delta\A
(\br)&=&-\sum_\q e^{-i\q\cdot \br} {\F(\q)\over q^2+ 1/
\lambda^2}\nn\\
{\bf F}(\q)&=&{4\pi\over c} \int d\br\,\, e^{i\q\cdot \br}\int d\br' \udK
(\br,\br') \A_0(z').\label{deltaAsoln}\eea

With the expressions (\ref{response}) and (\ref{deltaAsoln}) it is clearly
straightforward, if tedious, to evaluate the full spatial dependence of the
vector potential or the current around the impurity site. It is clear from
physical considerations that the current at any given point in space $\br$ will
depend both on the distance from the surface $z$ and on the distance from the
impurity $|\br-\R|$. In general, then, the currents and field ${
\B}=\nabla\times \A$ are functions of $x,y$ and $z$ everywhere. We postpone the
full evaluation to a subsequent study. At present we would like
 simply to obtain an estimate of the size of this effect, which we
 do by examining the perturbation of the component of the field
 along the initial applied field $\B(0)$ which we take to be along
 $y$.  The penetration depth is usually {\it defined}, even in
 cases where the decay of the fields is not  exponential,
 i.e. nonlocal electrodynamics, as
$\lambda_{tot} \equiv \int_0^\infty dz B_y/ B_y(0)$.
So for our current purposes we will simply adopt this definition and  evaluate
$B_y(x=y=0,z)$, i.e. at the transverse position of the impurity $\R=(0,0,z_0)$.
In this approximation we find \bea {\delta\lambda\over \lambda}\!\simeq
\!{\dA_x(0)\!+\!\int_0^\infty\!dz
\partial_x\!\dA_z\!+\!\lambda\!\partial_z\dA_x\!(0)\!-\!\lambda
\!\partial_x\!\dA_z(0)\over A_{0x}(0)}\eea
A careful examination of these terms and glance at Eq. (\ref{deltaAsoln}) shows
that the size of $\delta (1/\lambda^2)\simeq 2\delta\lambda/\lambda^3 $ is
simply set by $\udK(\p,\q)$ with $p,q\sim 1/\lambda$, as intuitively expected.
Expanding Eq. (\ref{dKmicro}) for 
$p,q\ll k_F$ and performing the
integration over energy $\xi_\k$ yields from the first two terms \bea \delta
K^{xx}(\p,\p)\simeq {ic\over 4\pi \lambda_0^2} T\sum_\omega {T_0\over {\cal
V}}\hskip -.1cm\int_0^{2\pi} \hskip -.3cm d\phi\,
{2\omega\Delta_\k^2(12\xi_1^2+\eta_\p^2)\over \xi_1^3(4 \xi_1^2+\eta_\p^2)^2},
\label{dKmicro2}\eea where $\xi_1\equiv
\sqrt{\omega^2+\Delta_\k^2}$, 
$T_0= (\pi N_0)^{-1}G_0/(c^2-G_0^2)$, with
$G_0=(1/2 \pi N_0){\rm Tr}\sum_\k \uG^0_{\k} =(2i\omega
/\pi\xi_1N_0) {\cal K}(\Delta_0/ \xi_1)$, 
is the $\tau_0$ component of the $T$-matrix, and $c$ is the
cotangent of the $s$-wave scattering phase shift 
($c=0$ corresponds to infinitely strong scattering).  ${\cal K}$ is
the complete elliptic integral of the first kind. Note that the
third term of (\ref{dKmicro}) contains two factors of $T_0^2$ and
therefore formally of order $1/N^2$; we therefore neglect it in
its effect on long-wavelength properties like $\delta\lambda$,
although it will contribute to local properties.
\begin{figure}[tb]
\begin{center}
\leavevmode
\includegraphics[width=.9\columnwidth]{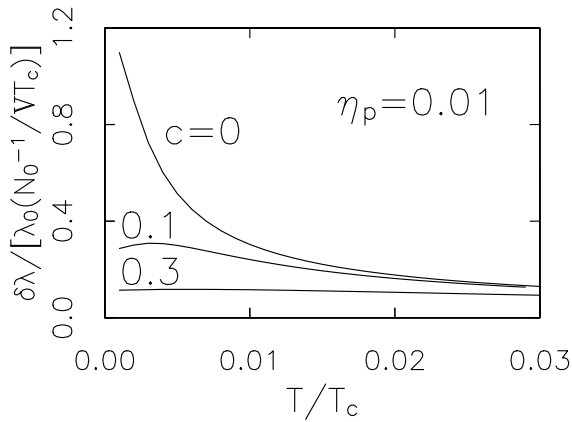}
\caption{Normalized change in penetration depth $\delta
\lambda/[\lambda_0(N_0^{-1}/{\cal V}T_c)]$ due to single isolated impurity vs.
normalized temperature $T/T_c$ for three values of the cotangent of the
impurity phase shift $c=0,0.1,0.3$. For this plot we took $\eta_\p\simeq 0.01
T_c$. } \label{fig:fig1}
\end{center}
\end{figure}  
The quantity $\eta_\p\equiv v_\k\cdot \p\simeq v_F^z/\lambda$, where the last
approximation follows because the primary spatial gradients are perpendicular
to the surface with normal $\hat z$, and because we are interested in typical
$p\simeq 1/\lambda$. In Figure \ref{fig:fig1}, we show a numerical evaluation
of (\ref{dKmicro2}) for three different values of the impurity scattering phase
shift.  As the scattering approaches the unitarity limit, the perturbation to
the penetration depth is seen to diverge with decreasing temperature.

To get an analytical estimate of the upturn in the unitarity limit, we first
perform a nodal expansion of the order parameter, $\Delta_\k\simeq
2\Delta_0(\phi-\phi_n)$, where $\phi_n\equiv \pi/4$, $3\pi/4$, ..., leading for
$T\ll\Delta_0$ to the result  \bea \delta K^{xx}(\p,\p)&\simeq& 
-{c\over
4\pi\lambda_0^2} { N_0^{-1}\over 2\pi^2 T{\cal V}}{1\over n_0
\log{2\Delta_0\over  \pi n_0 T}}\label{Kestimate},
 \eea
 where the infrared divergence in (\ref{dKmicro2}) was cut off
 by $\eta_\p$ or $c\Delta_0$, and $n_0$ is the appropriate minimum Matsubara index $n_0\equiv
 {\rm max} (\eta_\p/\pi T,c\Delta_0/\pi T,1)$.  For a   (001) surface in  the cuprates,
 $\eta_\p\simeq(v_F^z/v_F^\perp)(\xi_0/\lambda_0) \Delta_0 \simeq
 10^{-3}-10^{-4}$, so we get a divergence which goes as $\delta \lambda \sim
 \lambda_0 E_F/[NT\log(\Delta_0/T)]$.  The factor of $1/N$ arises
since (\ref{Kestimate}) is proportional to $N_0^{-1}/{\cal V}\sim a^3/{\cal
V}$, where $a$ is the lattice spacing. The estimate holds only for temperatures
down to $T_1^*\equiv {\rm max}(\eta_\p,c\Delta_0)$ at which the divergence is
 cut off, of order tenths of a Kelvin or less in the cuprates if $c$ is taken to be 0.

 {\it Finite density of scatterers.}  We would now like to extend the above estimate
for a single impurity to the case
 when a finite density of impurities is present.  As mentioned above,
 when brought into proximity
 these impurities interfere with one another via hybridization of quasiparticle bound state
 wave functions, leading eventually to an ``impurity band" and residual density of states
 at the Fermi level, at least in 3D\cite{pjhjltp}.  In this situation the SCTMA is
 the appropriate approximation, and a $T^2$ behavior in the penetration depth is found
 rather than an upturn.\cite{grossetal} The conditions for the
 formation of this band are subtle, and we do not address these questions here.

\begin{figure}[tb]
\begin{center}
\leavevmode
\includegraphics[width=1\columnwidth]{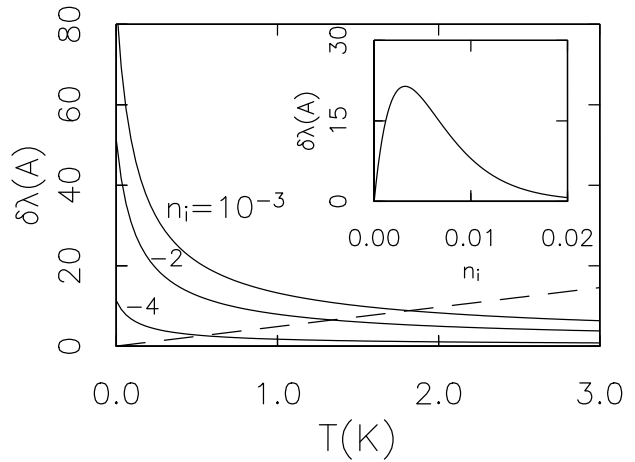}
\caption{Change in penetration depth $\delta \lambda_{tot}$(\AA) due to
isolated unitarity limit scatterers vs.  temperature $T(K)$ for three impurity
concentrations $n_i=10^{-2}$, $10^{-3}$, $10^{-4}$. $\Delta_0/T_c=2.14$,
$\eta_\p/T_c=0.001$, $T_c=100K$.  Dashed line: $T$-variation of pure
penetration depth $(\log 2) T/\Delta_0$. Inset: $\delta \lambda_{tot}$ vs.
$n_i$ at $T=1K$. } \label{fig:fig2}
\end{center}
\end{figure}

 Intuitively, it seems likely that impurities will act as independent
 sources of current distortion if they are sufficiently far from one another, but it is not so simple
 to specify what ``sufficiently far" means.
We note that both the
 distortion of the current distribution arising from $\delta K(\br,\br')$ {\it and} the
impurity bound state wave functions generically decay over lengths of a few
$\xi_0$. If the typical interimpurity spacing $\ell\equiv n_i^{-1/2}a$ (in 2D)
is much greater than $\xi_0$,  each impurity in a penetration length $\lambda$
contributes $\delta \lambda$ to the observed $\delta\lambda_{tot}$. On the
other hand, even if $\ell <<\xi_0$, there is a probability $P(R)\simeq \exp
{-R^2/2\ell^2} $ that an impurity will find itself with no other impurities
within a radius $R$, {\it if} all impurities are randomly distributed.  We
therefore expect that an {\it upper bound} to the observable penetration depth
upturn in the case of randomly distributed unitarity limit scatterers will be
\bea \delta \lambda_{tot}\simeq {n_i P(\xi_0) E_F\over 4\pi^2 \tilde T\log
(2\Delta_0/\pi \tilde T)} \lambda_0,\label{estimate2}\eea with $\tilde T = T +
T_1^*$. The estimate is an upper bound because we do not currently know how to
calculate accurately a geometric factor arising from the contribution of those
impurities which are located such that the tails of {\it nodal} quasiparticle
resonant states (which do  not decay as $e^{-\xi_0/r}$ but rather as $1/r$ at
$\omega=0$)\cite{balatsky,yashenkin} overlap. This factor will lead to
significant hybridization and smearing of the singularity at impurity densities
below that given by the criterion above. We nevertheless substitute some
numbers into (\ref{estimate2}) to check for plausibility, taking
$E_F/\Delta_0\simeq 10$, $\lambda =1500$\AA, and $\xi_0=25$\AA \hskip .1cm for
the cuprates. In Figure \ref{fig:fig2}, we plot the results of
(\ref{estimate2}) in laboratory units given these assumptions.
It is seen that, within this crude approximation, the size of the
upturn expected is not monotonic with impurity concentration, but
 at a temperature of 1K peaks around $n_i\simeq 10^{-3}$.
The position $T^*_2$ of the minimum in $\lambda_{tot}(T)$ will now be set by
comparing the impurity-induced change in $\lambda$ with that due to thermally
excited quasiparticles, $\delta\lambda(n_i,T^*_2)\sim T_2^*
\lambda_0/\Delta_0$; for  impurity concentrations of $n_i=10^{-3}-10^{-2}$,
this temperature will be of order 1-2K, as seen in Figure \ref{fig:fig2}.

  The size of
the increases predicted seem quite reasonable, and are of the same order of
magnitude as the upturns seen in the work of Bonn et al.\cite{bonnetal} and
Carrington et al.\cite{prozorov} Although the latter authors observed a
subtantial increase in the size of the signal when the proportion of (110)
surface in the given sample was maximized, indicative of an Andreev surface
bound state contribution, there was a substantial signal even in samples with
only (100) surfaces, where surface bound states should not exist. Although the
authors' suggestion that this result is due to (110) faceting in these samples
is 
possible, the mechanism we suggest may also be present.


While the  estimate (\ref{estimate2}) is based on the assumption of randomly
distributed impurities, it is important to recognize that in real systems
clustering will take place.  In this case even a relatively disordered system
may have significant numbers of isolated impurities or isolated atomic-scale
clusters which give rise to a singular current response.  In this case we might
expect significant sample-to-sample variation in $\delta \lambda$ in samples
with identical average concentration $n_i$.  This appears to be precisely the
effect found by Bonn et al. in their measurements on YBCO single crystals with
0.31\% Zn\cite{bonnetal}.

{\it Magnetic field.} Since the penetration depth upturn in the case considered
here is due to the large number of quasiparticle excitations near the nodal
directions, as is  the similar upturn in the case of Andreev surface states, we
might expect any physical effect which smears the gap nodes to cut off the
upturn. In particular, the orbital coupling to an applied magnetic field
(nonlinear electrodynamics) will suppress the upturn as it does in the Andreev
case\cite{barashetal}.  We need only add the Doppler shift of the quasiparticle
energy in any of the expressions above, $i\omega\rightarrow i\omega +\v_s \cdot
\k$, where $\v_s$ is the local superfluid velocity.  A typical shift
$v_sk_F\simeq (H/H_0)\Delta_0$, where $H$ is the applied field and $H_0
=3\Phi_0/(\pi^2\xi_0\lambda_0)$ is of the order of the thermodynamic critical
field, and 
the effect of the field may be included approximately by
generalizing $\tilde T\rightarrow T + T_1^* +(H/H_0)\Delta_0$ in
(\ref{estimate2}). We postpone a more quantitative study of the field
dependence to a later work.

{\it Conclusions.} 
We predict that a $d$-wave superconductor with
isolated strong nonmagnetic impurities should exhibit an upturn in the
penetration depth varying as $1/T \log T$ below a temperature scale set by the
disorder in the system and above one set by the bulk penetration depth or the
scattering phase shift.   This effect may have contributed to upturns already
observed in experiments where Andreev surface states, which produce a similar
effect, are not indicated\cite{bonnetal,prozorov}.   The physical origin of the
upturn  in both cases is  the depletion of the supercurrent at low $T$ in
low-energy bound states, but the  impurity-induced upturn  is independent of
surface geometry.  We argue that the magnitude of the effect should have a
characteristic dependence not only on the concentration of impurities, but on
the exact nature of the statistical distribution of impurities in the sample.
We emphasize that the estimates presented here are merely a crude plausibility
argument that if such upturns are observed at low temperatures, they could
arise from singular current response of local defects in the crystal, and that
it will be useful to study the dependence on disorder. Our work also allows in
principle an exact calculation of the local current flow around an impurity; it
will be interesting 
to see how the singular quasiparticle
currents are distributed in the neighborhood of the impurity site, and whether
these features can be detected by STM.

{\it Acknowledgements.} This work is supported by NSF
 DMR-9974396, by BMBF 13N6918/1, and by a grant from the A. v. Humboldt
foundation. The authors are grateful to Jochen Mannhart and the EP6 Lehrstuhl
at the University of Augsburg for hospitality during the completion of this
work, and to Russ Giannetta for a valuable communication.

\end{document}